\documentclass{optica-article}

\journal{opticajournal} 

\articletype{Research Article}

\usepackage{lineno}

\usepackage{graphicx}
\usepackage{dcolumn}
\usepackage{bm}

\usepackage{color}
\usepackage{xcolor}
\usepackage{soul}
\usepackage{upgreek}

\begin{document}


\title{Terahertz cavity magnon polaritons}

\author{
    T. Elijah Kritzell\authormark{1,2,*}, Andrey Baydin\authormark{1,3,*,$\dag$}, Fuyang Tay\authormark{1,2}, Rodolfo Rodriguez\authormark{4}, Hiroyuki Nojiri\authormark{5}, Henry O. Everitt\authormark{1,3,6}, Igor Barsukov\authormark{4}, and Junichiro Kono\authormark{1,3,7,8,$\dag\dag$}
}
\address{
    \authormark{1}Department of Electrical and Computer Engineering, Rice University, Houston, TX 77005, USA\\
    \authormark{2}Applied Physics Graduate Program, Smalley--Curl Institute, Rice University, Houston, TX 77005, USA\\
    \authormark{3}Smalley--Curl Institute, Rice University, Houston, TX 77005, USA\\
    \authormark{4}Physics and Astronomy, University of California, Riverside, CA 92521, USA\\
    \authormark{5}Institute for Materials Research, Tohoku University, Sendai 980-8577, Japan\\
    \authormark{6}DEVCOM Army Research Laboratory-South, Houston, TX 77005, USA\\
    \authormark{7}Department of Physics and Astronomy, Rice University, Houston, TX 77005, USA\\
    \authormark{8}Department of Material Science and NanoEngineering, Rice University, Houston, TX 77005, USA\\
    \authormark{*}The authors contributed equally to this work.\\
}
\email{\authormark{\dag}baydin@rice.edu
\authormark{\dag\dag}kono@rice.edu}


\begin{abstract*} 
Hybrid light--matter coupled states, or polaritons, in magnetic materials have attracted significant attention due to their potential for enabling novel applications in spintronics and quantum information processing. However, most studies to date have been carried out for ferromagnetic materials with magnon excitations at gigahertz frequencies. 
Here, we have investigated strong resonant photon--magnon coupling at frequencies above 1~terahertz for the first time in a prototypical room-temperature antiferromagnetic insulator, NiO, inside a Fabry--Pérot cavity. The cavity was formed by the crystal itself when it was thinned down to an optimized thickness. 
By using terahertz time-domain spectroscopy in high magnetic fields up to 25\,T, we swept the magnon frequency through Fabry--Pérot cavity modes and observed photon--magnon anticrossing behavior, demonstrating clear vacuum Rabi splittings exceeding the polariton linewidths. 
These results show that NiO is a promising platform for exploring antiferromagnetic spintronics and cavity magnonics in the terahertz frequency range. 
\end{abstract*}



\section{Introduction}
Strong resonant coupling of photons and magnons 
is expected to lead to novel device applications for spintronics, quantum transduction, and quantum information processing~\cite{LiEtAl2022, WangJAPP2020, LachanceApex2019, HarderBook}. 
Strong photon--magnon coupling was first demonstrated in ferrimagnets and ferromagnets placed inside microwave cavities~\cite{HueblEtAl2013PRL,TabuchiEtAl2014PRL,ZhangEtAl2014PRL}. Since then, many different types of applications of strong photon--magnon coupling have been proposed~\cite{BhoiKim2020SSP}. However, theoretical and experimental studies on strong photon--magnon coupling in antiferromagnets (AFMs) have been limited although, compared with ferromagnetic materials, AFMs are insensitive to external magnetic perturbations, do not generate stray fields, and exhibit ultrafast dynamics~\cite{RezendeEtAl2019JoAP}. Due to the high frequencies of antiferromagnetic resonances, AFM-based devices can operate in the terahertz (THz) frequency range, while ferromagnets are restricted to GHz frequencies. 

To date, several studies have shown strong photon--magnon coupling, demonstrating vacuum Rabi splittings larger than the linewidths of the lower and upper polariton peaks, for antiferromagnetic resonances~\cite{GrishuninEtAl2018AP,SivarajahEtAl2019JAP,BialekEtAl2020PRB,ShiEtAl2020arxiv,BialekEtAl2021PRA,MetzgerEtAl2022APL,BaydinEtAl2023PRR,BlankEtAl2023APL}. However, the studied AFMs either possessed sub-THz magnon frequencies~\cite{GrishuninEtAl2018AP,BialekEtAl2020PRB,BialekEtAl2021PRA,BlankEtAl2023APL}, or required low temperatures~\cite{GrishuninEtAl2018AP,ShiEtAl2020arxiv,MetzgerEtAl2022APL,BlankEtAl2023APL}, high magnetic fields~\cite{BaydinEtAl2023PRR}, or hybridization with other collective excitations~\cite{SivarajahEtAl2019JAP}. Most of these demonstrations have been made in rare-earth orthoferrites~\cite{BalbashovEtAl1995HFPiMM,LiEtAl2022}, in which antiferromagnetic resonances have low damping. Among other AFMs, NiO is a prototypical easy-plane AFM~\cite{SpinSuperFluid} with a high N\'{e}el temperature of 525\,K; this material has recently gained particular interest due to a series of studies showing current-induced magnetic switching~\cite{klaeuiNiONano, LuqiaoNiOPRL, KlaeuiNiOPRL}. The Gilbert damping factor of NiO's antiferromagnetic resonance at $\sim$\,1\,THz can be as low as $2.1\times10^{-4}$~\cite{KampfrathEtAl2011NP}, which is comparable to that of yttrium iron garnet and is thus suitable for strong photon--magnon coupling realization~\cite{YuanWang2017APL}. 

Here, we investigated two single-crystal wafers of NiO, each 10\,mm in diameter and 491\,$\upmu$m in thickness, cut along the (110) and (111) planes, respectively, via THz time-domain spectroscopy (THz-TDS) in high magnetic fields up to 25\,T~\cite{NoeEtAl2016OE,BaydinEtAl2021FO,TayEtAl2022JPSJ}. 
Our total spectral range of 0.1--1.5\,THz covered four of the previously reported magnon modes. We showed that only two modes, at 0.13 and 1\,THz, are infrared-active and have two-lobe symmetry with respect to the polarization of the THz excitation beam for both the (110)- and (111)-oriented samples.
The magnetic field dependence of the frequency of the 1\,THz mode was superlinear, consistent with the two-sublattice model~\cite{RezendeEtAl2019JoAP}. 
Furthermore, we observed anticrossing between the 1\,THz magnon mode and several Fabry--Pérot (FP) cavity photonic modes by sweeping the magnetic field. These results demonstrate strong photon--magnon coupling at room temperature, and therefore, are promising for developing devices for THz spintronics and quantum information processing applications. 

\section{Materials and Methods}
\begin{figure}
    \centering
    \includegraphics{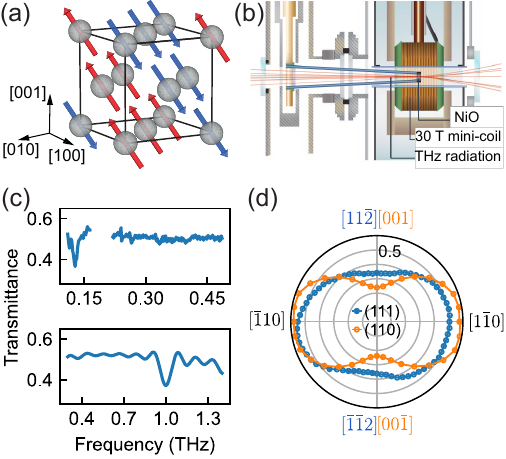}
    \caption{
    (a)~Ni$^{2+}$ ion arrangement in the NiO crystal, showing their spin orientations. Oxygen ions are not shown. 
    (b)~Schematic diagram of the THz time-domain spectroscopy setup in high magnetic fields (RAMBO)~\cite{NoeEtAl2016OE}.
    (c)~Transmittance spectra for the (111) NiO crystal obtained using a THz ellipsometer~\cite{HofmannEtAl2010RoSI} [Top] and THz-TDS [bottom].
    (d)~Excitation anisotropy for the 1\,THz magnon mode as a function of the magnetic field component of the THz beam, for the (111)- and (110)-cut NiO crystals. 
    }
    \label{fig:schematic}
\end{figure}

NiO is an easy-plane AFM. Below the N\'{e}el temperature, the spins of the Ni$^{2+}$ ions order parallel within each of the \{111\} planes, as shown in Fig.~\ref{fig:schematic}(a). The spins can align along one of the three axes: $\langle 11\bar{2} \rangle, \langle 1\bar{2}1 \rangle, \langle \bar{2}11 \rangle$ due to magnetic anisotropy. In addition to theoretical studies~\cite{Stevens1972JPCSSP,CottamAwang1979JPCSSP}, there have been several experimental reports on antiferromagnetic magnons in NiO studied by time-resolved Faraday rotation, THz-TDS, Raman, and Brillouin scattering measurements. In total, five magnon modes have been observed by different techniques. Two high-frequency antiferromagnetic resonances have been confirmed by Raman scattering~\cite{Lockwood1992AMO,LockwoodEtAl1992JoMaMM,GrimsditchEtAl1998PRB}, while three lower-frequency antiferromagnetic resonances have been confirmed by Brillouin scattering~\cite{GrimsditchEtAl1994JoMaMM,MilanoEtAl2004PRL}. Far-infrared optical absorption measurements have also confirmed the existence of the 1\,THz mode~\cite{Kondoh1960JPSJ,Tinkham1962JoAP,SieversTinkham1963PR,DietzEtAl1971PRB, PishkoEtAl2003JoAP}. Most recently, Kohmoto \textit{et al}.\ have employed optical pump-probe spectroscopy experiments with a circularly polarized pump beam and a Faraday rotation probe beam to observe three magnon modes at about 1.3, 1.1, and 0.1\,THz, respectively. They also found a narrow absorption line at 1.1\,THz and a broad absorption line at 0.5\,THz using THz-TDS down to low temperatures~\cite{KohmotoEtAl2018JIMTW}.

In order to explain the existence of these five magnon modes, an eight-sublattice model has been proposed~\cite{MilanoEtAl2004PRL}. The magnetic field dependence of 
some of these modes has been investigated by Brillouin scattering~\cite{MilanoGrimsditch2010PRB} and time-resolved Faraday rotation measurements~\cite{WangEtAl2018APL} to refine the parameters used in the eight-sublattice model further. The Brillouin scattering data were limited to the three lowest-frequency modes and magnetic fields up to 7\,T, while time-resolved Faraday rotation data provided magnetic field-induced evolution of the 1\,THz mode up to 10\,T in a limited spectral range. Despite these insights into the spin dynamics of NiO, various discrepancies between the model and experimental data~\cite{MilanoGrimsditch2010PRB,WangEtAl2018APL} have been reported, thus calling upon an extension of the experimental data set. 

To probe the magnetic field dependence of the magnons and photon--magnon coupling in NiO, we utilized a single-shot THz-TDS system, coupled with a table-top pulsed magnet with a peak field strength of 30\,T -- the Rice Advanced Magnet with Broadband Optics (RAMBO)~\cite{NoeEtAl2016OE,BaydinEtAl2021FO,TayEtAl2022JPSJ} (see Fig.~\ref{fig:schematic}b). In THz-TDS, we used a 775\,nm optical pulse from an amplified Ti:Sapphire laser (150\,fs, 1\,kHz, 0.8\,mJ, Clark-MXR, Inc., CPA-2001) for generation and detection of THz radiation. THz pulses were generated via optical rectification in LiNbO$_3$ and guided to the sample using parabolic mirrors. After going through the sample, the THz pulses were detected via electro-optic sampling in a ZnTe crystal. Single-shot detection of THz pulses was achieved using a reflective echelon and a fast CCD camera~\cite{NoeEtAl2016OE,BaydinEtAl2021FO,TayEtAl2022JPSJ}. The sample was placed on a cold sapphire rod that could reach temperatures down to 12\,K. The maximum magnetic field used was 25\,T. Transmittance spectra below 0.3\,THz were obtained using a THz ellipsometer, as detailed in Ref.\,\cite{HofmannEtAl2010RoSI}. Because an in-depth understanding of antiferromagnetic spin dynamics is required for employing AFMs in spin-torque devices and other spintronic applications~\cite{AFMSpinInjection, MagnonInteraction, AFMFMTransitions, gonccalves2018oscillatory}, we first focus on the behavior of the antiferromagnetic magnon modes and then discuss the observed phenomenon of strong photon--magnon coupling.

\section{Results and Discussion}
Figure~\ref{fig:schematic}(c) shows transmittance spectra for the (111) NiO sample, collected without the magnet, on a large optical aperture to provide the largest spectral bandwidth. Two magnon modes can be clearly identified at 0.13\,THz [Top] (taken using THz ellipsometer) and 1\,THz [Bottom], respectively. According to these spectra (blue traces), these two modes are the only infrared-active antiferromagnetic resonances. Note that this result is in disagreement with an earlier THz-TDS study of NiO~\cite{KohmotoEtAl2018JIMTW}, where a broad absorption line at around 0.5\,THz was reported. Although this feature may be absent from our measurements because our sample is thinner ($\approx0.5$\,mm vs. $\approx5$\,mm), it must be noted that such broad features may in principle be an artifact arising from setup limitations. 
We note that in a more recent study utilizing a cw-THz spectrometer~\cite{MoriyamaEtAl2019PRM}, similarly, only the 1\,THz magnon mode was observed in the range between 0.4\,THz and 1.25\,THz. 

Figure~\ref{fig:schematic}(d) shows the measured excitation (absorption) anisotropy for the 1\,THz magnon mode for the (111) and (110) NiO crystals. The data were collected at room temperature using a motorized rotation stage. For both crystals, we find similar two-fold excitation anisotropy. By looking at the magnetic structure [see Fig.~\ref{fig:schematic}(a)], we can expect two-fold symmetry for the (110)-cut crystal and six-fold symmetry for (111)-cut crystal. The six-fold symmetry expectation can arise because of three equivalent axes of spin alignment, $\langle 11\bar{2} \rangle$, $\langle 1\bar{2}1 \rangle$, $\langle \bar{2}11 \rangle$. While, for the (110)-cut sample, our data shows clear expected two-fold symmetry, the (111)-cut sample exhibits less anisotropy but still a two-fold symmetry. A similar two-fold anisotropy was also observed in both crystals for the 0.13-THz mode.

NiO has in total 12 orientational domains: four possible $\langle 111 \rangle$ stacking directions define twin domains, and per each twin domain there are three $\langle 11\bar{2} \rangle$ spin domains~\cite{HutchingsSamuelsen1972PRB,NakahigashiEtAl1975JPSJ,SangerEtAl2006PRB}. According to a recent second harmonic generation (SHG) study~\cite{SangerEtAl2006PRB}, both coherent and incoherent random assemblies of all twelve orientational domains result in a six-fold SHG signal anisotropy. This leads us to conclude that our sample preferentially consists of fewer than twelve orientational domains. 

\begin{figure}
    \centering
    \includegraphics{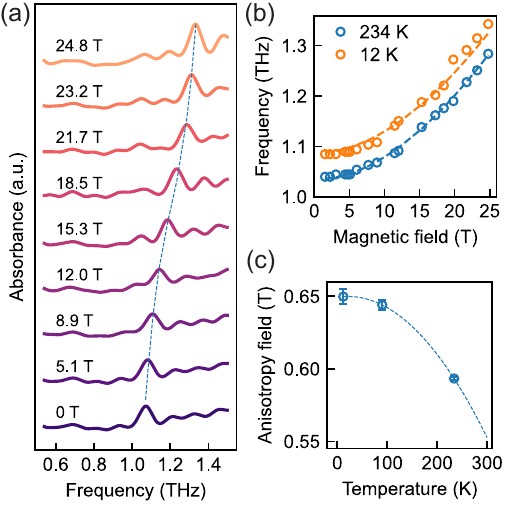}
    \caption{
    (a) Absorbance of (111) NiO as a function of magnetic field at 90\,K for several magnetic fields. All traces except the 0\,T trace are vertically offset for clarity. The dashed blue line indicates the main feature.
    (b) Magnon frequency as a function of magnetic field at two temperatures. Open circles are from the experimental data, and the super-linear dashed lines are the fits to Eq.~(\ref{eq:magnon_frequency}).
    (c) Effective anisotropy as a function of temperature, as derived from fitting data in (b). The dashed line is a guide to the eye.
    }
    \label{fig:abs_coef}
\end{figure}

The data in high magnetic fields has a limited spectral bandwidth compared to the 0\,T data, and therefore, we only report the evolution of the 1-THz magnon. Figure~\ref{fig:abs_coef}(a) shows measured absorbance spectra for the (111) NiO sample at different magnetic fields. The 1-THz magnon mode blue-shifts as the magnetic field is increased to about 25\,T, which is indicated by the blue dashed line. 
Figure~\ref{fig:abs_coef}(b) summarizes the magnetic field dependence of the frequency of this mode, where dashed lines are the fits, as explained below.

The literature values of the effective fields in NiO are $\mu_0 H_\mathrm{E} = 968.4$\,T for the exchange field and $\mu_0 H_\mathrm{HA} = 0.635$\,T and $\mu_0 H_\mathrm{EA} = 0.011$\,T  for the hard-axis ($\langle 111 \rangle$) and easy-axis ($\langle 11\bar{2} \rangle$) anisotropies, respectively, while the Land\'e g-factor $\mathfrak{g} = 2.18$~\cite{RezendeEtAl2019JoAP}. 
Due to the fact that we observe only two infrared-active modes, for simplicity it is sufficient to utilize a two-sublattice model. To assess the temperature dependence of magnetic anisotropy in our sample, we estimate the effective anisotropy of the samples using the equation for an easy-plane biaxial AFM frequency~\cite{BorovikRomanovEtAl1985JETP, SafinEtAl2022M}:

\begin{equation}\label{eq:magnon_frequency}
    \omega/\gamma_r = [2 H_\mathrm{E} H_\mathrm{A} + H_0^2]^{1/2},
\end{equation}
where $H_\mathrm{A} = H_\mathrm{HA} + H_\mathrm{EA}$ and the external magnetic field $H_0$ is perpendicular to the easy plane. Because $H_\mathrm{HA} \gg H_\mathrm{EA}$, the effective anisotropy is basically the hard-axis anisotropy, i.e., $H_\mathrm{A} \approx H_\mathrm{HA}$. Here, $\gamma_r$ is the gyromagnetic ratio $\mathfrak{g} \mu_\mathrm{B} / \hbar$, where $\mathfrak{g}$ is the Land\'e g-factor and $\mu_\mathrm{B}$ is the Bohr magneton. 

We fit the magnetic field-dependent magnon frequencies by Equation~(\ref{eq:magnon_frequency}), while fixing the parameters given above and adjusting the effective anisotropy field $H_\mathrm{A}$ as the fitting parameter. The effective anisotropy field obtained from this fitting procedure is presented in Fig.~\ref{fig:abs_coef}(c) as a function of temperature. The values of the hard-axis anisotropy are close to those reported in the literature~\cite{RezendeEtAl2019JoAP}. The temperature dependence of the anisotropy can be approximated by using either the square root of the classical molecular field or the molecular field including biquadratic exchange, as detailed in Ref.~\cite{KohmotoEtAl2018JIMTW}. 

As mentioned above, an eight-sublattice model has been developed and employed recently~\cite{WangEtAl2018APL} to describe magnetic field-dynamics of the NiO magnon observed by Brillouin scattering~\cite{MilanoGrimsditch2010PRB} and time-resolved Faraday rotation~\cite{WangEtAl2018APL} experiments. The eight-sublattice model predicts all five modes observed experimentally~\cite{MilanoEtAl2004PRL}, and the magnetic field dependence of two upper modes (1.15\,THz and 1.29\,THz at 0\,K) has been calculated~\cite{WangEtAl2018APL}. The 1.29-THz mode was shown to change with the magnetic field while the 1.15-THz mode was shown to be magnetic field independent. Wang~\textit{et al.}~\cite{WangEtAl2018APL} assigned these two modes to the experimentally observed 1-THz mode, which splits into two as the magnetic field is increased at low temperatures. The authors correlated experimentally observed modes and theoretically predicated modes by their magnetic field dependence, but their frequencies at 0\,T were not consistent. It is also important to note that the authors could not resolve the higher-frequency mode at 1.29\,THz at low temperatures (previously observed in Raman scattering) due to their frequency bandwidth~\cite{WangEtAl2018APL}. A more recent report~\cite{OhmichiEtAl2022JPSJ} has also commented on inapplicability of this model to explain their magnetic field dependent data. Thus, we add our experimental data for the ongoing discussion on the behavior of magnon modes in NiO.

\begin{figure}
    \centering
    \includegraphics{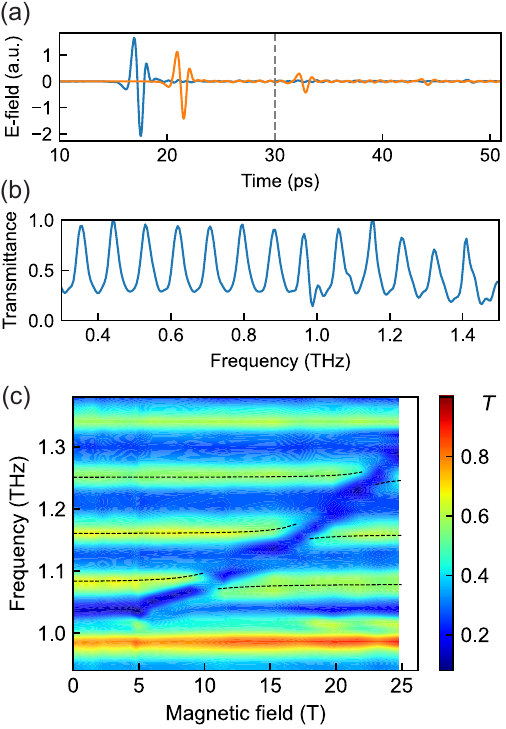}
    \caption{
    (a)~THz electric field transmitted through the sample (orange) and an empty aperture (blue). The spectra, including those in Fig.~\ref{fig:schematic}(c) and Fig.~\ref{fig:abs_coef}(a), were obtained through Fourier transform after cutting the time-domain signal before the back reflection (indicated by the gray dashed line). 
    (b)~Transmittance spectrum obtained from the full THz time-domain trace, at zero magnetic field and room temperature. 
    (c)~Color map of transmittance as a function of frequency and magnetic field at 234\,K, showing anticrossing features between the magnon mode and Fabry--Pérot cavity modes.
    }
    \label{fig:coupling}
\end{figure}

Now, we focus on strong coupling between FP cavity modes and the high-frequency ($\approx 1$\,THz) magnon mode. 
A FP cavity represents the simplest cavity structure formed by two interfaces. Instead of placing materials under study in between two mirrors, a slab of material in vacuum will act as a FP cavity itself. Thus, when the light wavelength is comparable to the thickness of such a slab, FP fringes (i.e., FP cavity modes) can be easily observed; (see \ref{fig:coupling}(b)). In conventional THz-TDS data analysis, where the FP contributions are intentionally removed from spectra by cutting the time-domain waveform before back-reflection signals appear~\cite{NeuSchmuttenmaer2018JoAP}; e.g., the absorbance shown in Fig.~\ref{fig:abs_coef}(a) was calculated after such a process. Here, however, we calculate transmittance by using the full range of the time domain signal that includes several back reflections; see Fig.~\ref{fig:coupling}(a). The resulting transmittance is shown in Fig.~\ref{fig:coupling}(b). The spacing between the FP cavity modes is 89\,GHz, and their full width at half maximum (FWHM) is 28\,GHz. For the thickness of the present sample, 491\,$\upmu$m, the magnon and FP cavity modes are detuned at zero magnetic field and room temperature. 

Figure~\ref{fig:coupling}(c) shows a transmittance color map as a function of temperature and magnetic field. As the magnon mode frequency increases with the magnetic field, it goes through three FP cavity modes in the measured frequency range. In total, three anticrossings are observed, which indicate formation of cavity magnon polaritons and, therefore, strong photon--magnon coupling. Note that the frequency of the FP cavity mode can be adjusted by changing the sample thickness, allowing for strong photon-magnon coupling to occur at 0\,T.
The photon-magnon coupling strength $g$ can be estimated from the eigenfrequencies of the quantum Langevin equations~\cite{GaoNatPhot2018}:
\begin{equation}
    \omega_{\pm} = \frac{1}{2}(\omega + \omega_\mathrm{cav}) - \frac{i (\gamma + \kappa)}{4} \pm \sqrt{ g^2 + \left( \frac{i (\kappa - \gamma)}{4} + \frac{(\omega - \omega_\mathrm{cav})}{2} \right)^2},
    \label{eq:freqs}
\end{equation}
where $\omega_{\pm}$ are the lower and upper polariton frequencies, $\omega$ is the magnon frequency, $\omega_\mathrm{cav}$ is the FP cavity mode frequency, and $\gamma$ and $\kappa$ are the magnon and cavity decay rates, respectively. 

The dashed black lines in Fig.~\ref{fig:coupling} were calculated by Eq.~(\ref{eq:freqs}) with parameter values of $g=0.014$\,THz, $\gamma = 0.015$\,THz, and $\kappa = 0.028$\,THz.  The value of $\kappa$ is the FWHM of the FP cavity modes. The value of $\gamma$ can be estimated from the magnon linewidth in free space.  However, since our measured linewidths were instrument limited (25\,GHz), we used the literature value of 15\,GHz~\cite{MoriyamaEtAl2019PRM}.  With these parameter values, we get the cooperativity $C = 4g^2/\gamma\kappa = 1.87$, which satisfies the condition of strong coupling $C>1$~\cite{PeracaEtAl2020SaS}.  Note that these values of $g$ and $C$ quantitatively describe the anticrossing behavior of all FP cavity modes shown in Fig.~\ref{fig:coupling}(c) and were the same for all temperatures where THz measurements were made (12\,K, 90\,K, 234\,K).

\section{Conclusion}
We studied the temperature and magnetic field dependence of THz antiferromagnetic resonances in NiO and their coupling to FP cavity modes, where the sample itself acted as a cavity. Two infrared-active magnon modes (0.13 and 1\,THz) were observed in a broad frequency range, from 0.1\,THz to 1.6\,THz, with a two-lobed excitation anisotropy with respect to the polarization of the THz pulse. At high magnetic fields, we also found a secondary mode centered at about 1\,THz, which remains nearly independent of the magnetic field. 
Examining the interactions between the 1\,THz magnon mode and FP cavity modes, we observed strong photon--magnon coupling with cooperativity $C = 1.87$ and normalized coupling strength $g/\omega_0 = 0.014$ for the 12$^\mathrm{th}$ FP mode. The realization of strong photon--magnon coupling in AFMs in the THz frequency range can bring spintronics and information processing applications to the THz operating regime. 

\begin{backmatter}
\bmsection{Funding}
The U.S.\ National Science Foundation (ECCS-1810541), 
the Gordon and Betty Moore Foundation (A23-0150-001), the W.\ M.\ Keck Foundation (995764), the Robert A.\ Welch Foundation (C-1509), and the U.S.\ Army Research Office (W911NF2110157)

\bmsection{Acknowledgments}
R.R.\ and I.B.\ acknowledge support by the National Science Foundation under Grant No.~NSF-ECCS-1810541. J.K.\ acknowledges support from the Gordon and Betty Moore Foundation (Grant No.\ A23-0150-001), the W.\ M.\ Keck Foundation (Grant No.\ 995764), the U.S. Army Research Office (Award No. W911NF2110157), and the Robert A.\ Welch Foundation (Grant No.\ C-1509). 

\bmsection{Disclosures}
The authors declare no conflicts of interest.

\bmsection{Data Availability Statement}
Data underlying the results presented in this paper are not publicly available at this time but may be obtained from the authors upon reasonable request.

\end{backmatter}

\bibliography{2-references}

\end{document}